\documentclass{mem}
\usepackage{natbib}\usepackage{txfonts}\usepackage{balance}
\usepackage{graphicx}
\usepackage[breaklinks]{hyperref}
\usepackage{etoolbox}
\makeatletter
\patchcmd\@combinedblfloats{\box\@outputbox}{\unvbox\@outputbox}{}{\errmessage{\noexpand patch failed}}
\makeatother
\pdfoutput=1
\begin{document}
\def\teff{$T\rm_{eff }$}
\def\kms{$\mathrm {km s}^{-1}$}
\newcommand{\eprint}{}

\bibpunct{(}{)}{;}{a}{}{,} 


\title{
Constraining the properties of neutron-star matter with observations
}

   \subtitle{}

\author{
Eemeli Annala\inst{1},
Tyler Gorda\inst{2},
Aleksi Kurkela\inst{3},
Joonas N\"attil\"a\inst{4}
\and Aleksi Vuorinen\inst{1},
          }

\institute{
Department of Physics and Helsinki Institute of Physics, P.O. Box 64, FI-00014 University of Helsinki, Finland
\and Department of Physics, University of Virginia, Charlottesville, Virginia 22904-4714, USA
\and Theoretical Physics Department, CERN, Geneva, Switzerland and Faculty of Science and Technology, University of Stavanger, 4036 Stavanger, Norway
\and Nordita, KTH Royal Institute of Technology and Stockholm University, SE-10691 Stockholm, Sweden\\
\email{aleksi.vuorinen@helsinki.fi}
}

\authorrunning{Annala et al.}

\titlerunning{Constraining the NS-matter EoS}

\abstract{
In this conference-proceedings contribution, we review recent advances in placing model-independent constraints on the properties of cold and dense QCD matter inside neutron stars. In addition to introducing new bounds for the Equation of State, we explain how these results may be used to make robust statements about the physical phase of strongly interacting matter in the centers of neutron stars of different masses. Our findings indicate that the existence of quark-matter cores inside massive neutron stars appears to be a very common feature of the allowed Equations of State, and should not be considered an exotic or unlikely scenario.
}
\maketitle{}

\section{Introduction}

The cores of neutron stars (NSs) contain the densest matter in our Universe, and probe the phase diagram of Quantum Chromodynamics (QCD) in a regime that has never been directly experimentally studied \citep{Brambilla:2014jmp}. Interestingly, this dense and cold corner of the phase diagram is also by far the hardest to study theoretically. This is due to the infamous Sign Problem of lattice QCD that prohibits a numerical Monte-Carlo determination of the bulk thermodynamic properties of QCD matter already at moderate densities \citep{deForcrand:2010ys}. For this reason, the accuracy to which we understand the behavior of dense nuclear matter at densities exceeding the nuclear saturation density $n_s\approx 0.16/$fm$^3$ is rather limited (see, e.g.,\ \citealt{Tews:2012fj} for state-of-the-art results), while it is only at tens of saturation densities that the asymptotic freedom of QCD allows a treatment of the system using the machinery of perturbative field theory \citep{Laine:2016hma}. Considering that it is exactly this density interval that witnesses the deconfinement transition, where hadrons are replaced by quarks and gluons as the effective degrees of freedom in the system, it becomes clear that the cores of NSs form an extremely exciting laboratory for nuclear and particle physics. 

On the observational front, the past decade has been a remarkable period, marked by the discovery of two-solar-mass NSs \cite{Demorest:2010bx,Antoniadis:2013pzd}, the detection of Gravitational Waves (GW) from a NS merger \citep{TheLIGOScientific:2017qsa}, as well as simultaneous NS mass-radius (MR) measurements of increasing precision (see, e.g.,\ \citealt{Nattila:2015jra,Nattila:2017wtj}). 
Apart from settling a host of astrophysical puzzles, these advances have allowed model-independent studies of the collective properties of NS matter --- most importantly its Equation of State (EoS) (see, e.g.,\ \citealt{Annala:2017llu,Most:2018hfd}). Typically, such analyses utilize the robustly known low-and high-density limits of the EoS as well as commonly accepted astrophysical observations to construct a family of allowed NS matter EoSs. The gold standard in the field has become to interpolate the EoS between the low- and high-density limits using various basis functions, which include most prominently piecewise-defined polytropes and Chebyshev polynomials, the latter pioneered by Lindblom (see, e.g.,\  \citealt{Lindblom:2010bb}). 

While the above research avenue has set stringent limits for the NS matter EoS, limiting its uncertainty to a fraction of what it used to be pre-2010s, a closely related and very important question has received little attention until very recently. Namely, the EoS analyses carried out so far have been by construction blind to the identity of matter inside NS cores, i.e.,\ whether matter is in a hadronic or deconfined (quark matter) phase. This is, of course, only natural, considering that the motivation to interpolate the EoS over the intermediate-density regime has been to avoid dealing with the complicated dynamics near the deconfinement transition, where no first-principles microphysical calculations are possible. 

Despite the above reservations, it is not out of the question that with increasingly precise astrophysical constraints, the properties of NS matter could be determined to such a precision that firm, model-independent statements could be made about the physical phase of matter inside NS cores. Whether this is the case already is clearly a very interesting question, and one that has been investigated very recently in \citep{Annala:2019puf} with the help of a novel interpolation scheme for the EoS where the starting point is a linear interpolation of the speed of sound $c_s^2 = \mathrm{d} p/\mathrm{d}\epsilon$ of NS matter. The remainder of the conference-proceedings contribution at hand is devoted to reviewing this study, which produced very interesting results.

Our presentation is organized as follows. As any model-independent determination of the presence of quark matter (QM) inside NSs must be based on a careful analysis of the family of viable EoSs, Sec.~2 is devoted to explaining how the state-of-the-art NS EoS family of \citep{Annala:2019puf} was constructed. After this, we move on to analyzing the properties of matter at low and high densities in Sec.~3, and in particular asking the question of how matter in the centers of $1.44M_\odot$ stars, i.e.,\ typical pulsars, differs from that inside maximally massive stable NSs. Finally, in Sec.~4 we end our presentation with a brief discussion of the astrophysical implications of our findings.

\section{Constraining the NS-matter EoS}

The procedure used in studies such as \citep{Annala:2017llu,Most:2018hfd,Annala:2019puf} to generate model-independent families of NS-matter EoSs, covering in principle all possible physically allowed 
behaviors, is conceptually very straightforward:
\begin{enumerate}
 \item Take the low- and high-density limits of the EoS from state-of-the-art calculations in nuclear and particle theory. Below $n=1.1n_s$, one typically uses the Chiral Effective Theory (CET) computation of \citep{Tews:2012fj}, and above approx.~$40n_s$ the perturbative QCD (pQCD) calculations of \citep{Kurkela:2009gj,Gorda:2018gpy}.
 \item Introduce a set of interpolating functions (piecewise-defined  polytropes, Chebyshev polynomials, piecewise linear interpolations of the speed of sound squared,...) to parameterize the EoS in the intermediate-density region.
 \item Generate large ensembles of EoSs by varying the free parameters in the interpolating functions, making sure that the resulting EoSs respect different physical consistency conditions, such as the subluminality of the speed of sound.
\item Constrain the obtained (typically large, at least ${\mathcal O}(10^5)$) EoS family  by demanding that the resulting NSs agree with robust astrophysical observations, such as the existence of $2M_\odot$ NSs and the tidal-deformability limits of LIGO/Virgo.
\end{enumerate}

\begin{figure}[t]
\resizebox{\hsize}{!}{\includegraphics[clip=true]{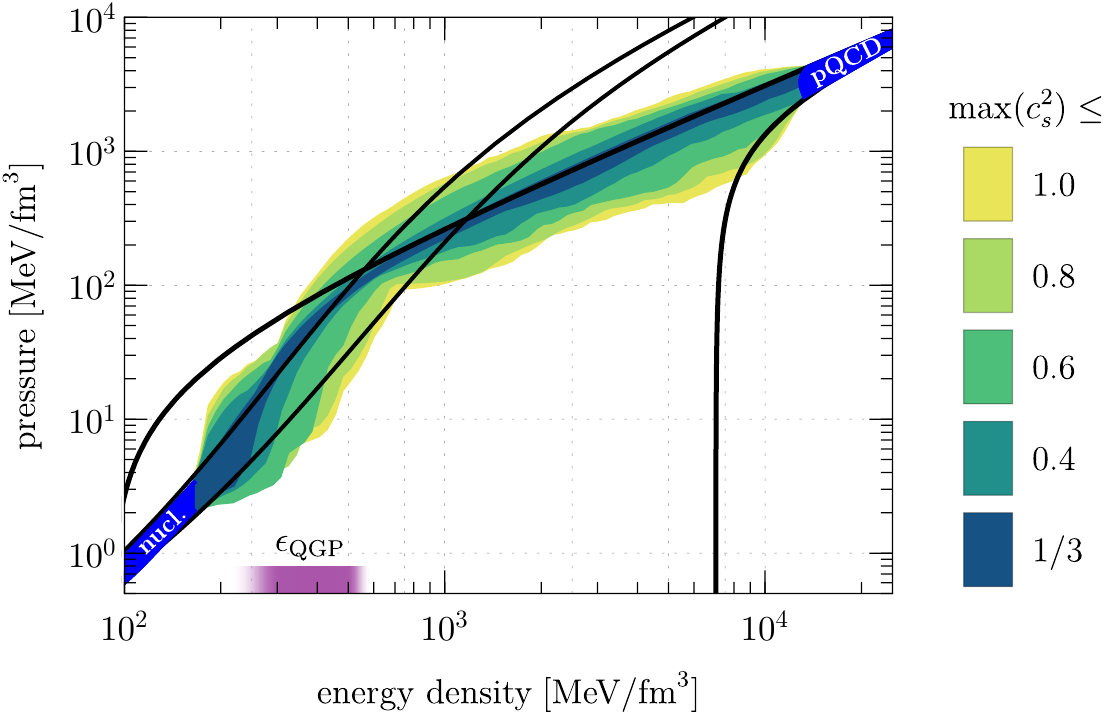}}
\caption{
\footnotesize
The band of allowed NS-matter EoSs constructed with the speed-of-sound interpolation method of \citep{Annala:2019puf}, which allows setting various upper limits for $c_s^2$, indicated with the color coding. The symbol $\epsilon_{\rm QGP}$ refers to the energy densities, where the deconfinement transition takes place in high-temperature QGP, while the black solid lines represent extrapolations of the theoretical low- and high-density EoSs beyond their regimes of validity.
}
\label{fig1}
\end{figure}

\begin{figure}[t]
\resizebox{\hsize}{!}{\includegraphics[clip=true]{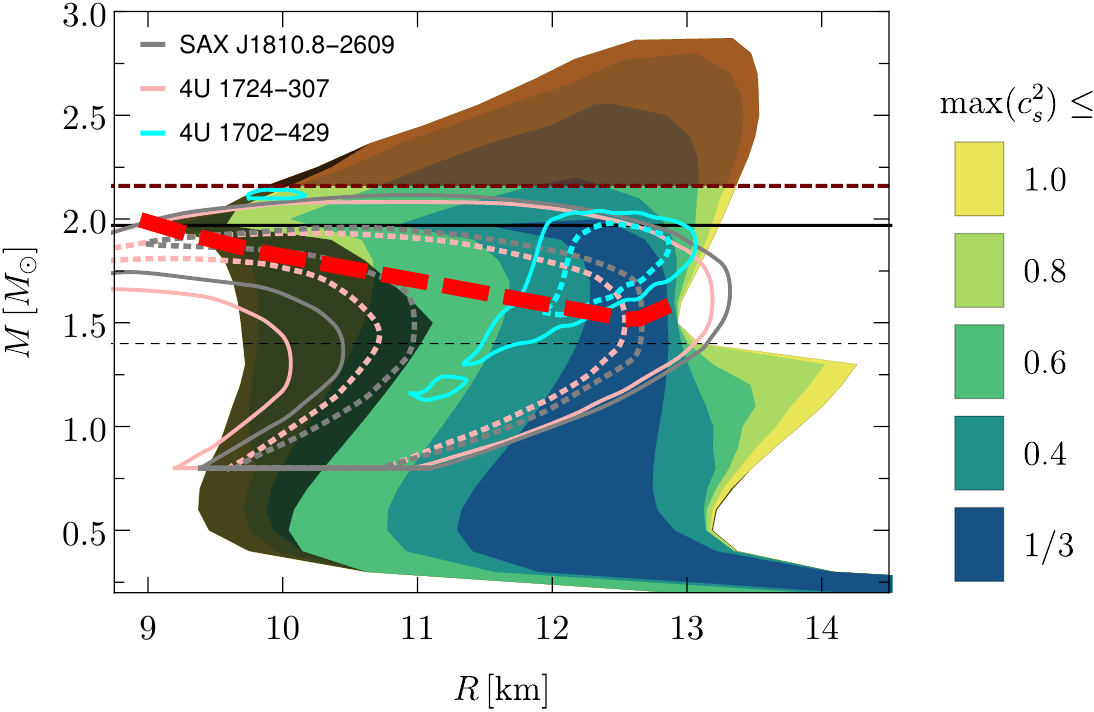}}
\caption{
\footnotesize
A mass-radius plot constructed with the EoS ensemble displayed in Fig.~\ref{fig1}. The dark regions on the left and on the top are excluded by constraints based on the EM counterpart of GW170817, in particular $\tilde \Lambda>300$ \citep{Radice:2018ozg} and $M_{\rm max}<2.16M_\odot$  \citep{Rezzolla:2017aly}. Below the red dashed line, there are no NSs with QM cores. 
}
\label{fig2}
\end{figure}

For the case of the most recent interpolation study of \citep{Annala:2019puf}, the result of the above exercise is displayed in Figs.~\ref{fig1} and \ref{fig2}. In the EoS plot, we also show extrapolations of the theoretical low- and high-density EoSs beyond their respective regimes of validity \citep{Gandolfi:2011xu,Kurkela:2009gj}, and in the MR plot the results of several recent simultaneous MR measurements \citep{Nattila:2015jra,Nattila:2017wtj}. In both cases, we display the results in a form in which the color coding reveals how the results would change were we 
to impose an upper limit for the speed of sound $c_s$ in dense QCD matter. Besides demonstrating that the edges of the distributions are composed of the highest-$c_s$ EoSs, this result shows that it is indeed possible to have the speed of sound consistently stay below the conformal bound $c_s^2=1/3$. This is a very interesting result, as there is a strong theoretical reason to expect that the bound 
not be violated by a sizable amount: there are no known physical systems where $c_s$ would exceed $1/\sqrt{3}$, and even theoretically the bound has only been violated in somewhat artificial setups (see also \citep{Cherman:2009tw,Bedaque:2014sqa,Hoyos:2016cob,BitaghsirFadafan:2018uzs,Ishii:2019gta}). 

What is particularly noteworthy about the EoS band obtained, and becomes more pronounced when nontrivial cuts are made on the speed of sound, is that the result appears to be composed of two nearly linear parts, joined at a ``kink'' around an energy density of order 500 MeV/fm$^3$.
This result strongly suggests a qualitative change in the material properties of the medium around this energy density, which happens to nearly coincide with one where hot quark-gluon plasma (QGP) is produced in heavy-ion collisions. Whether this very suggestive interpretation survives after a case-by-case analysis of the individual EoSs forming the band is, however, not obvious. Similarly, where the central energy densities of NSs of various masses are located in the EoS plot of Fig.~\ref{fig1}, is clearly an interesting topic to study.

\section{Existence of quark-matter cores} 

\begin{figure*}[t]
$\;\;\;\;\;\;\;\;\;\;\;\;\;\;\;\;\;\;\;\;\;\;\;\;\;\;\;\;\;\;\;$\resizebox{0.6\hsize}{!}{\includegraphics[clip=true]{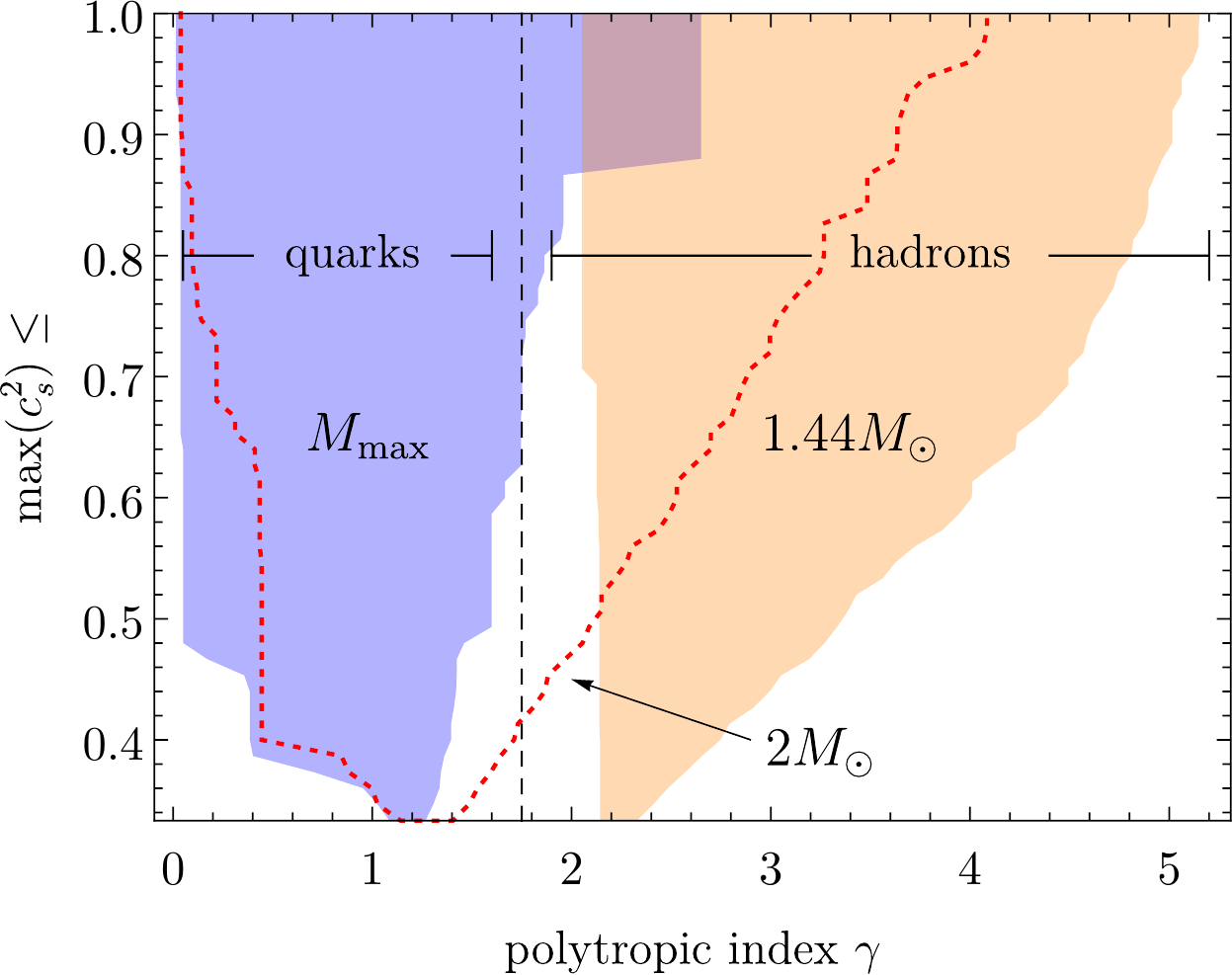}}
\caption{\footnotesize
The distribution of polytropic indices $\gamma$ at the centers of $1.44M_\odot$ and maximal-mass NSs, classified in terms of the maximal value the speed of sound reaches in the corresponding EoSs. The dashed red curve corresponds to the case of $2M_\odot$ stars, i.e.,\ the most massive NSs observed to date.
}
\label{fig3}
\end{figure*}

To reliably investigate the presence of QM in the cores of NSs, the procedure one needs to follow is again conceptually simple, but potentially nontrivial to carry out. For every individual EoS in our ensemble, one needs to determine the location of the deconfinement transition --- which can furthermore be either a first-order one or a crossover --- and then solve the Tolman-Oppenheimer-Volkov (TOV) equations to find out whether the stellar cores reach (and exceed) this density. For this, one clearly needs a quantitative, robust and easily automatable criterion for determining the onset of QM in a given EoS, as the only feasible way to go through all the EoSs is through automation.

As can be observed already from the CET and pQCD EoS curves shown in Fig.~\ref{fig1}, a physical feature that differentiates QM from hadronic matter (HM) is its stiffness. This can be quantified via the polytropic index $\gamma \equiv \mathrm{d}(\log p)/\mathrm{d}(\log \epsilon)$,  which takes values close to unity in high-density QM and values well above $2$ in low-density HM. This quantity is moreover very straightforward to evaluate, and so lends itself to an efficient classification of EoSs. To this end, in Fig.~\ref{fig3} we display the distributions of polytropic indices in the centers of 1.44$M_\odot$ and maximal-mass stars obtained for \emph{all} allowed EoSs in our ensemble, and moreover show the results assuming various upper limits for $c_s^2$. A number of very interesting observations can be made from here:
\begin{itemize}
\item Taking the value $\gamma=1.75$ as the dividing line between HM and QM, we see that the typical pulsar stars of 1.44$M_\odot$ mass are always composed of HM alone, but the maximal-mass stars predominantly have quark cores.
\item If one imposes nontrivial upper limits for $c_s$, the division between the two distributions becomes clearer with decreasing max$(c_s^2)$. For instance, if we choose to only look at EoSs for which $c_s^2<0.7$, then \emph{all maximal-mass stars contain QM cores}.
\item Inspecting the EoSs that lead to no QM cores in maximal-mass stars, one observes that it is always a
first-order phase transition that destabilizes the star. In other words, if we were to find out, e.g.,\ from theoretical studies that the deconfinement transition is of 
a crossover type even at high density, we would immediately know that QM exists inside some physical NSs.
\end{itemize}

\begin{figure}[t]
\resizebox{\hsize}{!}{\includegraphics[clip=true]{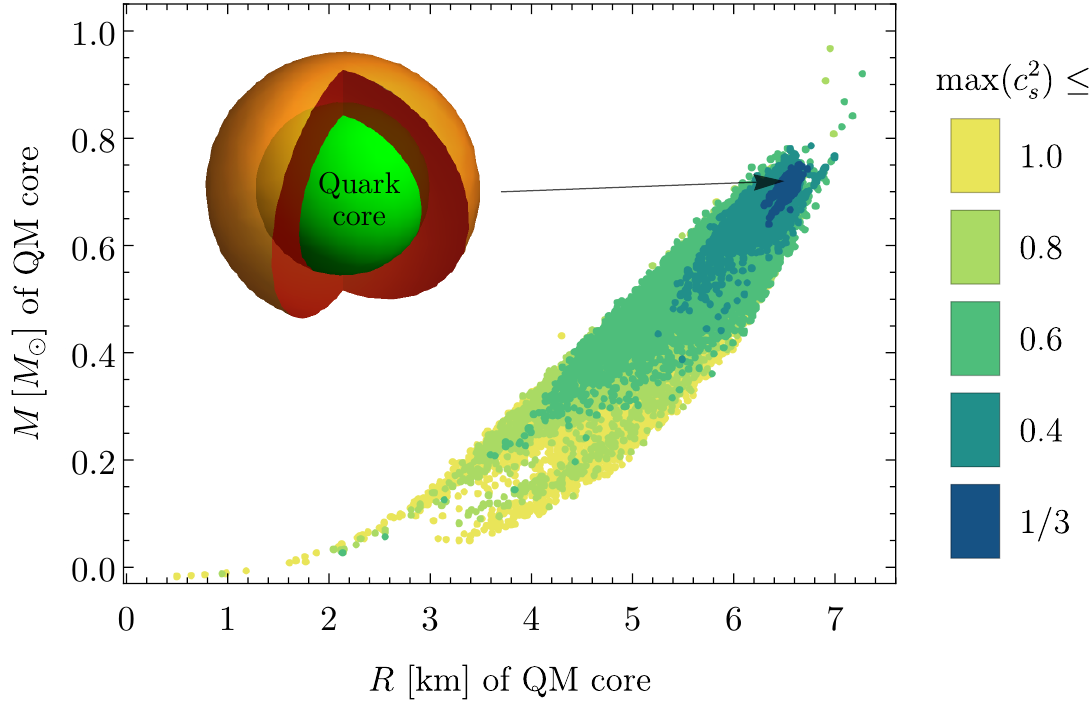}}
\caption{
\footnotesize
The radii and masses of the QM cores of maximal-mass NSs obtained with our EoS ensemble. For maximal speed-of-sound values close to the conformal limit $c_s^2=1/3$, we find large QM cores.
}
\label{fig4}
\end{figure}

In Fig.~\ref{fig4}, we next display the amount of QM found inside maximal-mass stars, discovering once again a strong correlation between the maximal speed of sound and the presence of QM inside stellar cores. Should the maximal value of $c_s^2$ remain below 0.5, all EoSs consistent with known observational bounds lead to very sizable QM cores, exceeding 0.25$M_\odot$ in mass. At the other end of the spectrum, the tail of the distribution continuing towards the origin is composed of EoSs with large maximal speeds of sound, and contains at the origin the configurations for which no QM exists at all. In this subset of EoSs, the maximal value of $c_s^2$ is always at least 0.7, and the minimal latent heat of the transition 130 MeV/fm$^3$.

\begin{figure}[t]
\resizebox{\hsize}{!}{\includegraphics[clip=true]{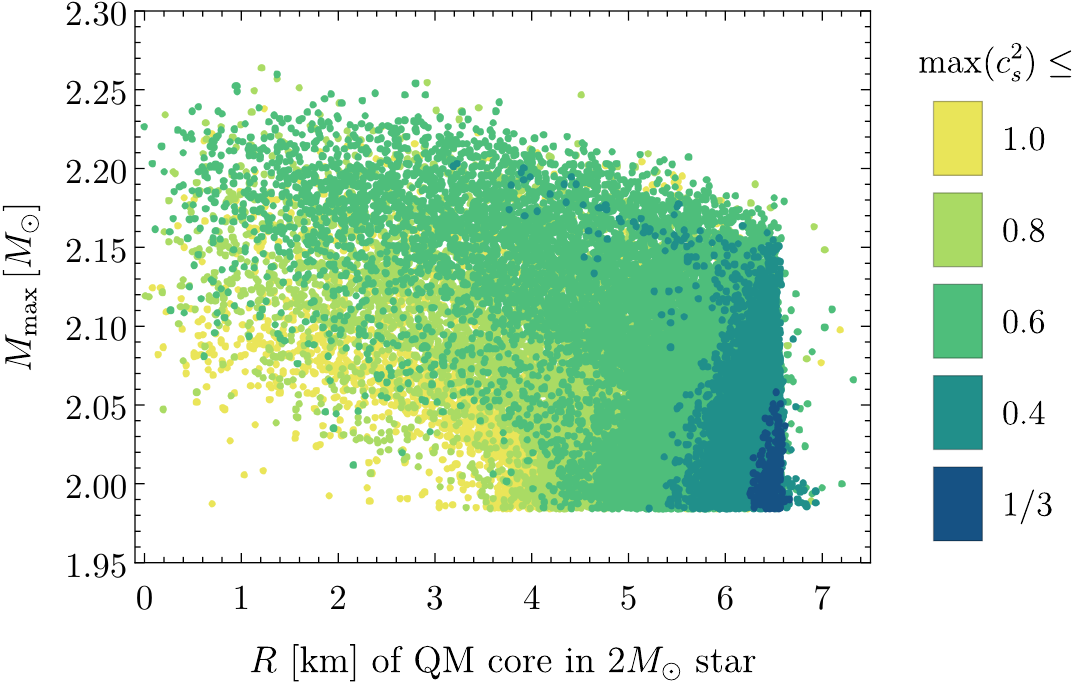}}
\caption{
\footnotesize
The radius of the QM core in two-solar-mass NSs vs.~the maximal mass of NSs built with the same EoS. Subconformal EoSs are seen to lead to large QM cores and low values of 
$M_{\rm max}$.
}
\label{fig5}
\end{figure}

An interesting further question is clearly the fate of QM inside the most massive NSs observed to date, i.e.,\ J1614$-$2230 and J0348$+$0432, with $M \approx 2 M_\odot$ 
\citep{Demorest:2010bx,Antoniadis:2013pzd}. This is studied in Fig.~\ref{fig5}, where we display the correlation between two quantities: the radius of the QM core in 2$M_\odot$ stars (whenever nonzero) and the maximal mass of stable NSs corresponding to the same EoS. The result is interesting in at least two respects: first, we observe that if QM is present already in two-solar-mass stars, the maximal mass cannot become larger than approx.~2.25$M_\odot$, indicating that the softening associated with the emergence of a QM core has a destabilizing effect on the stars even in the absence of a first-order deconfinement transition. Second, for EoSs that (almost) respect the conformal bound $c_s^2\leq 1/3$, the two-solar-mass stars always contain large quark cores,  of order 6.5 km for subconformal EoSs.

\section{Conclusions}

The observational study of neutron-star properties offers a unique way to probe the behavior of baryonic matter at extreme densities. This perhaps counterintuitive finding hinges on two separate facts: that matter in the cores of NSs is compressed to densities greatly exceeding the nuclear-matter saturation density, and that the material properties of NS matter are encoded in the macroscopic properties of the stars via the TOV equations. Simply put, by making increasingly accurate measurements of the masses and radii of NSs, as well as closely related quantities obtainable, e.g.,\ via gravitational-wave studies of NS mergers, we may systematically constrain the Equation of State of QCD matter to such an extent that this quantity may one day be accurately pinpointed.

In the conference proceedings at hand, we have reviewed results from a very recent study \citep{Annala:2019puf}, where constraining the EoS of NS matter with observations has been taken one step further than previously thought possible. By utilizing a novel method for interpolating the NS-matter EoS between the known low- and high-density limits of the quantity, we have demonstrated that current NS observations already allow for drawing model-independent conclusions about the physical phase of matter inside NS cores. In short, our results indicate that the existence of quark-matter cores inside very massive NSs should be considered the standard scenario, not an exotic alternative. QM is altogether absent in NS cores only under very specific conditions, including both a strong first-order deconfinement transition and a very high speed of sound inside NS matter. Conversely, should the transition be a crossover or the speed of sound violate the conformal bound $c_s=1/\sqrt{3}$ only mildly, the quark cores in many NSs, including the already observed $2M_\odot$ stars, are bound to be sizable.

Our findings concerning the ubiquity of QM cores in massive NSs may have some very interesting observable consequences. In the context of NS mergers, it is tempting to speculate about shock waves reflecting from the QM-HM interface inside hypermassive NSs, in particular if there is a sizable difference between the speeds of sound in the two phases. Also, the presence of QM may lead to enhanced dissipation and subsequently increased damping during the ringdown phase, owing to a large effective bulk viscosity. Finally, the presence of sizable QM cores may enable an altogether new astrophysical field of study: NS seismology. Similarly to the Earth, there is a possibility of seismic waves (from, e.g., crust breaking) launching inwards to the center and reflecting back from the QM-HM interface. If this can lead to any observable effects like additional heating of the outer crust --- and subsequent emission of soft x-rays --- by shock waves bouncing back and forth remains to be investigated.

\begin{acknowledgements}
The work of EA, TG, and AV has been supported by the European Research Council, grant no.~725369. In addition, EA gratefully acknowledges support from the Finnish Cultural Foundation and TG from the U.S.~Department of Energy Grant No.~DE-SC0007984. This paper is associated with the preprint numbers HIP-2019-13/TH and NORDITA 2019-031.
\end{acknowledgements}

\bibliographystyle{aa}
\bibliography{vuorinen_integral}

\begin{thebibliography}{23}
\expandafter\ifx\csname natexlab\endcsname\relax\def\natexlab#1{#1}\fi

\bibitem[{Abbott {et~al.}(2017)}]{TheLIGOScientific:2017qsa}
Abbott, B. {et~al.} 2017, Phys. Rev. Lett., 119, 161101

\bibitem[{Annala {et~al.}(2019)Annala, Gorda, Kurkela, Nattila, \&
  Vuorinen}]{Annala:2019puf}
Annala, E., Gorda, T., Kurkela, A., Nattila, J., \& Vuorinen, A. 2019
  [\eprint[arXiv]{1903.09121}]

\bibitem[{Annala {et~al.}(2018)Annala, Gorda, Kurkela, \&
  Vuorinen}]{Annala:2017llu}
Annala, E., Gorda, T., Kurkela, A., \& Vuorinen, A. 2018, Phys. Rev. Lett.,
  120, 172703

\bibitem[{Antoniadis {et~al.}(2013)}]{Antoniadis:2013pzd}
Antoniadis, J. {et~al.} 2013, Science, 340, 6131

\bibitem[{Bedaque \& Steiner(2015)}]{Bedaque:2014sqa}
Bedaque, P. \& Steiner, A.~W. 2015, Phys. Rev. Lett., 114, 031103

\bibitem[{Bitaghsir~Fadafan {et~al.}(2018)Bitaghsir~Fadafan, Kazemian, \&
  Schmitt}]{BitaghsirFadafan:2018uzs}
Bitaghsir~Fadafan, K., Kazemian, F., \& Schmitt, A. 2018
  [\eprint[arXiv]{1811.08698}]

\bibitem[{Brambilla {et~al.}(2014)}]{Brambilla:2014jmp}
Brambilla, N. {et~al.} 2014, Eur. Phys. J., C74, 2981

\bibitem[{Cherman {et~al.}(2009)Cherman, Cohen, \& Nellore}]{Cherman:2009tw}
Cherman, A., Cohen, T.~D., \& Nellore, A. 2009, Phys. Rev., D80, 066003

\bibitem[{de~Forcrand(2009)}]{deForcrand:2010ys}
de~Forcrand, P. 2009, PoS, LAT2009, 010

\bibitem[{Demorest {et~al.}(2010)Demorest, Pennucci, Ransom, Roberts, \&
  Hessels}]{Demorest:2010bx}
Demorest, P., Pennucci, T., Ransom, S., Roberts, M., \& Hessels, J. 2010,
  Nature, 467, 1081

\bibitem[{Gandolfi {et~al.}(2012)Gandolfi, Carlson, \& Reddy}]{Gandolfi:2011xu}
Gandolfi, S., Carlson, J., \& Reddy, S. 2012, Phys. Rev., C85, 032801

\bibitem[{Gorda {et~al.}(2018)Gorda, Kurkela, Romatschke, Sappi, \&
  Vuorinen}]{Gorda:2018gpy}
Gorda, T., Kurkela, A., Romatschke, P., Sappi, M., \& Vuorinen, A. 2018, Phys.
  Rev. Lett., 121, 202701

\bibitem[{Hoyos {et~al.}(2016)Hoyos, Jokela, Rodriguez~Fernandez, \&
  Vuorinen}]{Hoyos:2016cob}
Hoyos, C., Jokela, N., Rodriguez~Fernandez, D., \& Vuorinen, A. 2016, Phys.
  Rev., D94, 106008

\bibitem[{Ishii {et~al.}(2019)Ishii, Järvinen, \& Nijs}]{Ishii:2019gta}
Ishii, T., Järvinen, M., \& Nijs, G. 2019 [\eprint[arXiv]{1903.06169}]

\bibitem[{Kurkela {et~al.}(2010)Kurkela, Romatschke, \&
  Vuorinen}]{Kurkela:2009gj}
Kurkela, A., Romatschke, P., \& Vuorinen, A. 2010, Phys. Rev., D81, 105021

\bibitem[{Laine \& Vuorinen(2016)}]{Laine:2016hma}
Laine, M. \& Vuorinen, A. 2016, Lect. Notes Phys., 925, pp.1

\bibitem[{Lindblom(2010)}]{Lindblom:2010bb}
Lindblom, L. 2010, Phys. Rev., D82, 103011

\bibitem[{Most {et~al.}(2018)Most, Weih, Rezzolla, \&
  Schaffner-Bielich}]{Most:2018hfd}
Most, E.~R., Weih, L.~R., Rezzolla, L., \& Schaffner-Bielich, J. 2018, Phys.
  Rev. Lett., 120, 261103

\bibitem[{Nattila {et~al.}(2017)Nattila, Miller, Steiner, Kajava, Suleimanov,
  \& Poutanen}]{Nattila:2017wtj}
Nattila, J., Miller, M.~C., Steiner, A.~W., {et~al.} 2017, Astron. Astrophys.,
  608, A31

\bibitem[{Nattila {et~al.}(2016)Nattila, Steiner, Kajava, Suleimanov, \&
  Poutanen}]{Nattila:2015jra}
Nattila, J., Steiner, A.~W., Kajava, J. J.~E., Suleimanov, V.~F., \& Poutanen,
  J. 2016, Astron. Astrophys., 591, A25

\bibitem[{Radice \& Dai(2018)}]{Radice:2018ozg}
Radice, D. \& Dai, L. 2018 [\eprint[arXiv]{1810.12917}]

\bibitem[{Rezzolla {et~al.}(2018)Rezzolla, Most, \& Weih}]{Rezzolla:2017aly}
Rezzolla, L., Most, E.~R., \& Weih, L.~R. 2018, Astrophys. J., 852, L25,
  [Astrophys. J. Lett.852,L25(2018)]

\bibitem[{Tews {et~al.}(2013)Tews, Kruger, Hebeler, \& Schwenk}]{Tews:2012fj}
Tews, I., Kruger, T., Hebeler, K., \& Schwenk, A. 2013, Phys. Rev. Lett., 110,
  032504

\end{thebibliography}

\end{document}